\documentstyle[12pt]{article} 
 \advance\oddsidemargin by -1in 
\advance\evensidemargin
 by -1in 
\advance\topmargin by -0.0in 
\makeindex

\newcommand\beq{\begin{equation}}
 \newcommand\eeq{\end{equation}} 
\newcommand\beqn{\begin{eqnarray}}
 \newcommand\eeqn{\end{eqnarray}} 
\newcommand{\doublespace} {
 \renewcommand{\baselinestretch} {1.6} 
\large\normalsize}

\begin{document}
\vspace*{0.5cm}
\hspace*{10cm}{\Large DFTT 56/95}\\
\hspace*{10.6cm}{\Large Torino, 1995}\\
\vspace*{2cm}
 
 \centerline{\huge \bf Color Transparency or}
\smallskip
\centerline{\huge {\bf the Standard Inelastic Shadowing?}\footnote
{Based on the  talks
 presented by B.~Kopeliovich at the Workshop 
{\it Color Coherence/Transparency
 Studies at CEBAF}, CEBAF, May 1995, and at 
the ELFE Workshop, Cambridge, July 1995}}
\vspace{.5cm}
\begin{center}
{\Large Boris~Kopeliovich}\footnote{On leave from 
 Joint Institute for
 Nuclear Research, Laboratory of Nuclear Problems,
\newline Dubna, 141980 Moscow Region, Russia.  E-mail:
 bzk@dxnhd1.mpi-hd.mpg.de}
 
\vspace{0.3cm}
 
 {\sl Max-Planck Institut f\"ur Kernphysik, Postfach 103980,
\newline 69029 Heidelberg, Germany}\\
 
\vspace{0.6cm}
 
 {\Large Jan Nemchik}\footnote{On leave from Institute of Experimental Physics
 SAV, Solovjevova 47, CS-04353 Kosice, Slovakia}
 
\vspace{0.3cm}
 
 {\sl Dipartimento di Fisica Teorica, Universit\`a di Torino\\ I-10125, Torino,
 Italy}
\end{center}

\vspace{1cm}
\doublespace
\begin{abstract}

The standard first-order inelastic correction (IC) well 
known in the pre-QCD era, causes a
rising $Q^2$-dependence of nuclear 
transparency in the quasielastic electron scattering,
$A(e,e'p)A'$, at moderate $Q^2$, similar to what 
is supposed to be the onset of color transparency (CT).
Although IC is a part of the whole
pattern of CT, it contains no explicit QCD dynamics.
Evaluation of this correction is based on experimental
data on diffraction dissociation and is 
independent of whether CP phenomenon exists or not.
The growth of nuclear transparency
is numerically comparable with
the expected signal of CT up to about $Q^2\approx 20\ GeV^2$.
Analogous effect in $A(p,2p)A'$ reaction is estimated as well.

\end{abstract}

\newpage
 
\doublespace

 Recent failure of the NE18 experiment at SLAC \cite{ne18} to observe the onset
 of color transparency (CT) in $(e,e'p)$ reaction
 has excited interest to the baseline for such a study. It was realized
 that even the Glauber model predictions have a substantial uncertainty.
Nevertheless,  a nearly
$Q^2$-independent nuclear
 transparency is expected in the Glauber approximation,  what makes it
 possible to single out the $Q^2$-dependent onset of CT effects \cite{m,b}.

We call Glauber eikonal approximation an approach disregarding
any off diagonal diffractive rescatterings of the ejectile, which itself
is supposed to be just a proton even in the primary interaction of the
electron with a bound proton.  

\medskip 

 It is known since Gribov's paper \cite{gribov}, that 
the  Glauber model should be corrected for inelastic shadowing at high energies.
The very existence and the numerical evaluations of the inelastic shadowing effect
was nicely confirmed by the high precision measurements of the total
cross sections of interaction of neutrons \cite{murthy} and neutral K-mesons
\cite{gsponer} with nuclei. Due to the inelastic corrections (IC) 
the total cross section 
turns out to be smaller, i.e. nuclear matter is more transparent, than is
expected in the Glauber approximation. Important for further discussion is
the rising energy dependence of nuclear transparency due to the growth of
IC.
An example is shown in fig.1. The data for $n-Pb$
total cross section as function of energy are compared 
with the 
Glauber approximation corrected or not for the
IC, evaluated in \cite{murthy,kk}.

\begin{figure}[tbh]
\includegraphics{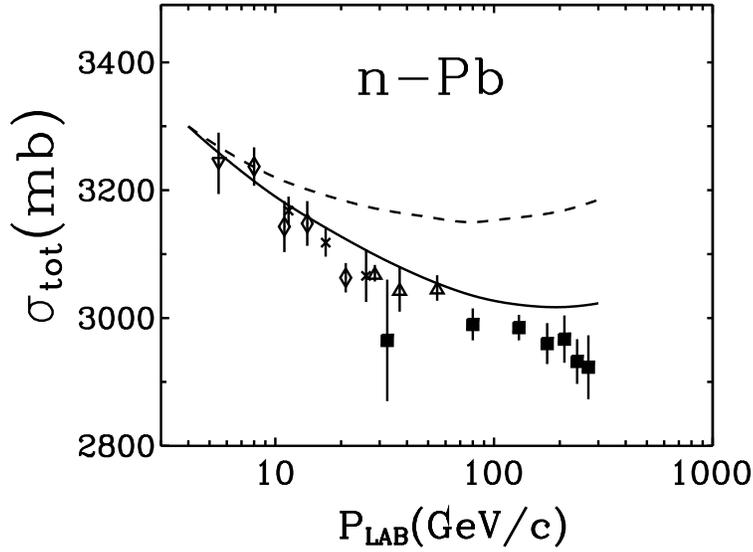}
\begin{center}
\vspace{8cm}
\parbox{13cm}
{\caption[Delta]
{Data on $n-Pb$ total cross
section (ref. \cite{murthy} and references therein). 
The dashed curve is corresponds to the Glauber approximation.
The solid line shows the effect of inclusion of the first order
IC to the total cross section 
as it is calculated in ref. \cite{murthy}}
\label{event}}
\end{center}
\end{figure}

According to these results one can expect that nuclear
matter becomes more transparent at higher $Q^2$ because
the ejectile energy correlates with $Q^2$, $\nu = Q^2/2m_N$.
Such a rising $Q^2$-dependent nuclear transparency
can imitate the CT effects which are expected to
manifest themselves as a monotonous 
growth of nuclear transparency with
$Q^2$. 

\medskip

At this point it worth reminding that CT 
is a particular case of Gribov's inelastic
shadowing, provided that QCD dynamics 
tunes many elastic and inelastic diffractive 
rescatterings to cancel each other in 
the final state interaction \cite{zkl,jm} at high $Q^2$. 
Therefore, one may 
think that there
is no sense in picking up only one IC
from many others, which all together
build up CT. However,
searching for CT effects one should first of all ask himself 
what happens
if CT phenomenon does not exist; for instance, if the ejectile 
in $(e,e'p)$ on a bound nucleon is not a small-size wave
packet, but is a normal
proton. Even in this case the IC shown schematically in fig. 2,  
makes the nuclear
matter more transparent than is predicted by the Glauber model.

\begin{figure}[tbh]
\includegraphics{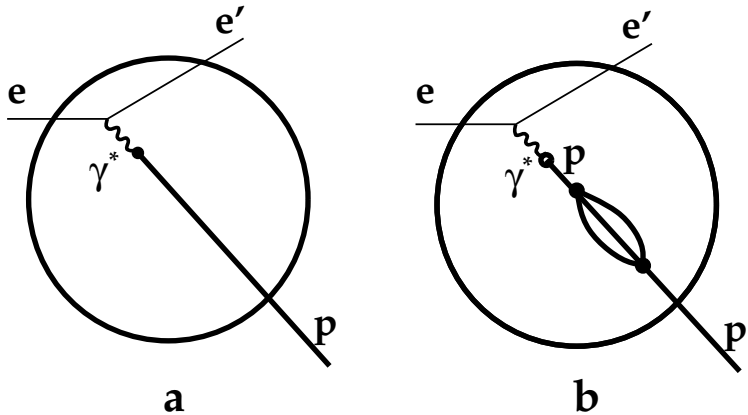}
\begin{center}
\vspace{6cm}
\parbox{13cm}
{\caption[Delta]
{Cartoon showing $A(e,e'p)A'$ reaction with eikonal elastic final state 
interactions (a) and with diffractive production of inelastic intermediate 
state (b)}
\label{event}}
\end{center}
\end{figure}

This first order IC corresponds to the diffractive 
production of inelastic intermediate states by the ejectile proton
while it propagates through the nucleus. 
  The proton waves with and without this
 correction interfere with each other, while the contributions from
different production points points add up incoherently 
because the momentum transfer 
in the $(e,e'p)$ reaction is too large. 
It is important that IC has a positive relative sign, provided that all
diffraction amplitudes are imaginary, 
as was demonstrated in \cite{kl,jk,fgms}.
The resulting nuclear transparency reads,
 
 \beqn 
Tr(Q^2) & = & \int d^2b \int_{-\infty}^{\infty}
 dz\rho(b,z)\ \exp[-\sigma^{NN}_{in} \int_z^{\infty} dz' \rho(b,z')]\times\nonumber\\ 
& & \left[1+4\pi \int dM^2 \frac{d\sigma}{dM^2dt}|_{t=0}\
 F^2_A(b,z,q_L)\right]^2
\label{2}
\eeqn
 
 Here $b$ and $z$ are the impact parameter and longitudinal coordinate of the
 bound proton which absorbs the virtual photon.  $\sigma^{NN}_{in}$ is
 inelastic $NN$ cross section. We assume that the $(e,e'p)$ 
cross section is integrated
over the transverse momentum of the ejectile proton relative to the photon direction. 
$T(b)=\int_{-\infty}^{\infty}dz \rho(b,z)$ is
 the nuclear thickness function.  $d\sigma/dM^2dt|_{t=0}$ is the forward cross
 section of single diffraction dissociation in $NN$ interaction.  $F_A(b,z,q_L)$
 is the nuclear longitudinal form factor \cite{kk},
 
 \beq 
F_A(b,z,q_L)=\int_{z}^{\infty}dz'\ \rho(b,z')e^{iz'q_L}\ ,
\label{3}
\eeq 
where $q_L=(M^2-m_N^2)/2\nu$ is the longitudinal momentum transfer in the
diffraction dissociation. This form factor leads to the $Q^2$-dependence of
 nuclear transparency.

Note that we assume in (\ref{2}) that the cross section is integrated over
the angle between the proton and the virtual photon momenta (see \cite{kolya3}).
 
One can find in \cite{murthy} a detailed calculation of IC
to the nuclear total cross section.
We use the same parameterization of the data on
$d\sigma/dM^2dt$ as in \cite{murthy} and the realistic nuclear density from
\cite{jager} to calculate expression (\ref{2}).  Following refs. 
\cite{kk,murthy,gsponer} we assume that
the inelastic intermediate states attenuate in nuclear medium 
with the same inelastic cross section as the proton. The predicted 
growth of nuclear transparency with $Q^2$ in $Pb(e,e'p)$ is compared in fig. 3
with what is expected to 
be a onset of CT \cite{kolya}. We use $\sigma^{NN}_{in} =
33\ mb$ in order to have the same transparency in the Glauber approximation
as in ref. \cite{kolya}. We see that 
these two mechanisms, one with and another one without CT dynamics
predict about the same magnitude of deviation from the Glauber model
up to about $Q^2\approx 20\ GeV^2$. It is especially difficult to disentangle
the real CT effects and the first-order IC
because of a substantial model-dependence 
of the theoretical predictions for CT. In order to 
detect reliably a signal of CT one needs $Q^2$ at least of a 
few tens of $GeV^2$. As different from CT,
 the growth of transparency provided by the first 
IC saturates at high $Q^2$ at quite a low level.

\begin{figure}[tbh]
\includegraphics{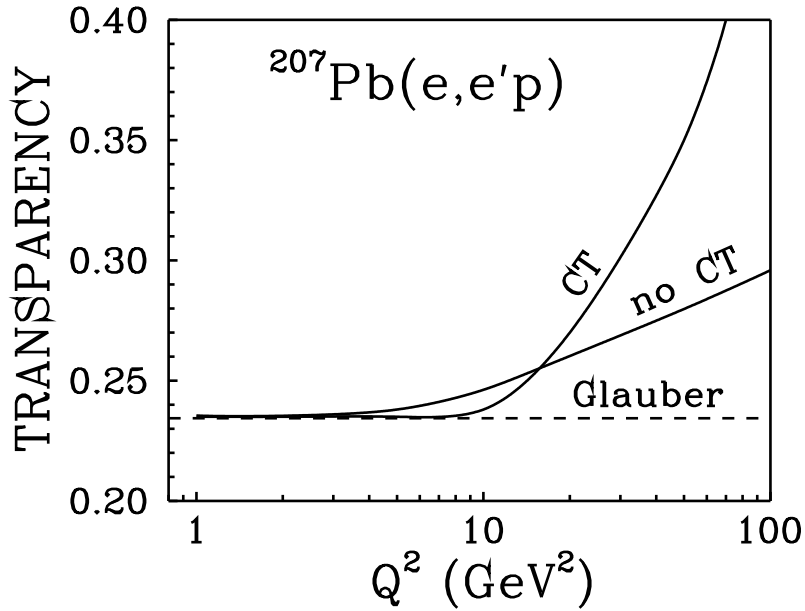}
\begin{center}
\vspace{9cm}
\parbox{13cm}
{\caption[Delta]
{Comparison of the Glauber model (dashed line) with the
model \cite{kolya} incorporating with CT (solid curve -- CT) 
and with our calculation of the first-order IC using eq.(\ref{2}) 
(solid curve
-- no CT)}
\label{event}}
\end{center}
\end{figure}

Our calculations are compared with the data from the
NE18 experiment at SLAC \cite{ne18} in fig. 4. In order to be more realistic
we use the data on $\sigma_{in}^{NN}$ from ref. \cite{compil} which exhibit
a decreasing energy-dependence at low energies (compare with \cite{fsz}).
Of course, more sophisticated calculations may consider the effects of Fermi 
motion \cite{jk,bbk,kolya2}, few-nucleon correlations 
\cite{benhar,kolya2,kohama,rinat,moniz},
accuracy of the closure approximation \cite{rj}, etc. We try to escape these 
complications to make the presentation simpler.
The relative contribution of IC is expected to be nearly
independent of the details of nuclear structure.

\begin{figure}[tbh]
\includegraphics{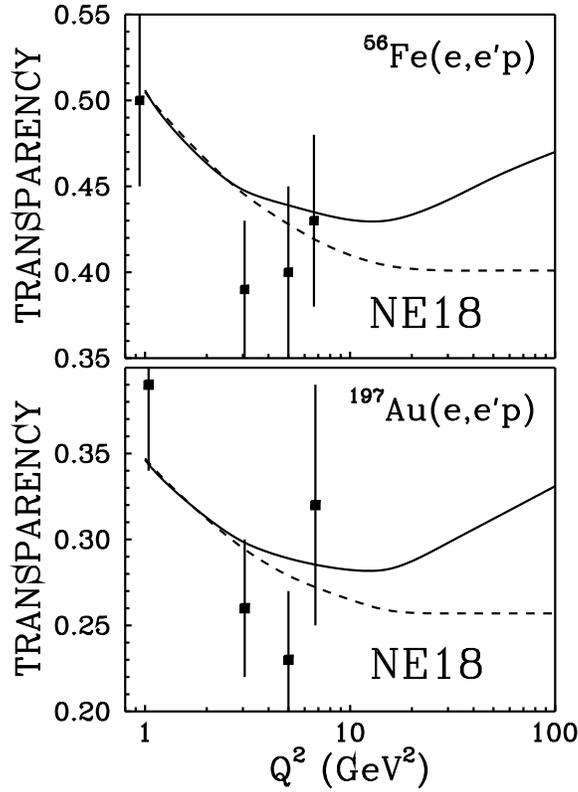}
\begin{center}
\vspace{11cm}
\parbox{13cm}
{\caption[Delta]
{Comparison of the Glauber model (dashed line) and of the results  
of our calculations of the standard first-order 
IC, eq. (\ref{2}), (solid line) 
with the data from the NE18 experiment \cite{ne18}}
\label{event}}
\end{center}
\end{figure}

Note that we predict a bigger effect of inelastic shadowing than that in
total hadron-nucleus cross sections \cite{murthy}. 
In the latter case case it is a correction to the small exponential term
in the elastic amplitude which is subtracted from unity, 
while in the present case we deal with a net
transparency effect.
 
Note that the same problems are inherent in the quasielastic 
hadron scattering off nuclei, for instance, 
$A(p,2p)A'$. We expect a growth of nuclear transparency 
due to the standard IC in this 
reaction as well. The effect of the 
first IC was 
taken into account in all three proton legs and
evaluated using the formula analogous to eq. 
(\ref{2}) and the realistic energy-dependent inelastic nucleon
cross section,

 \beqn
& & Tr(p,2p) = \int d^2b \int_{-\infty}^{\infty}
 dz\rho(b,z)\ \exp\left [-2\sigma^{NN}_{in} \int_z^{\infty} dz' \rho(b,z')\right ]
\times\nonumber\\
& & \exp \left [-\sigma^{NN}_{in} \int_{-\infty}^z dz' \rho(b,z')\right ]
\left[1+4\pi \int dM^2 \frac{d\sigma}{dM^2dt}|_{t=0}\
\left | \int_z^{\infty} dz' \rho(b,z) e^{2iz'q_L}\right |^2\right]^4\times\nonumber\\
& & \left[1+4\pi \int dM^2 \frac{d\sigma}{dM^2dt}|_{t=0}\ 
\left | \int_{-\infty}^z dz' \rho(b,z') e^{iz'q_L}\right |^2\right]^2
\label{4}
\eeqn

The results for lead are plotted in fig. 5 in comparison with the data from the BNL experiment
\cite{bnl}. Of course we do not pretend to fit to the data, which, we believe, 
is difficult to 
explain within any known realistic model. Fig.5 demonstrates the size 
of the effect due to the first-order IC.

\begin{figure}[tbh]
\includegraphics{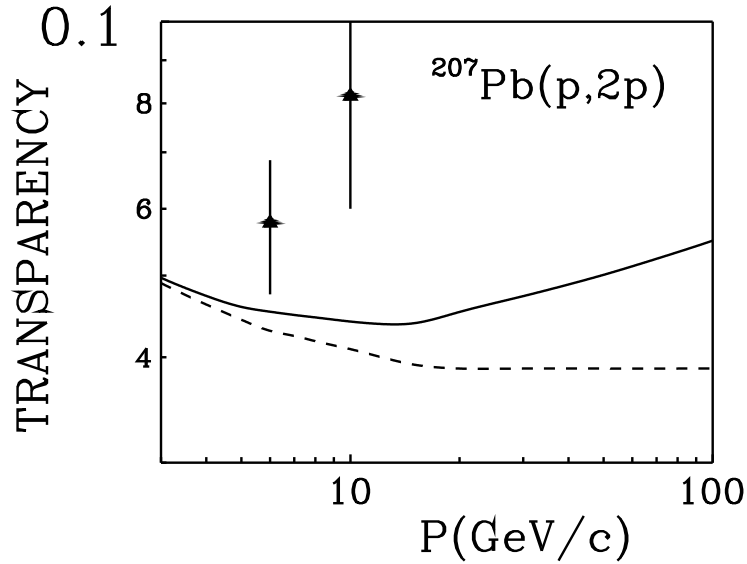}
\begin{center}
\vspace{8cm}
\parbox{13cm}
{\caption[Delta]
{The data from the Brookhaven experiment ref. \cite{bnl}
vs the Glauber model prediction (dashed curve)
and including the first order IC. eq. (\ref{4}) (solid curve).
We took into account the energy dependence of $\sigma_{in}^{pp}$}
\label{event}}
\end{center}
\end{figure}

 \medskip

To conclude, we estimated the first-order IC, which 
causes a growth of nuclear transparency with $Q^2$ 
in quasielastic scattering of electrons
and hadrons off nuclei and can imitate
the expected onset of CT up to $Q^2\sim 20\ GeV^2$.
In fact, this correction grows with the ejectile/projectile 
energy, which correlates with $Q^2$.
 The evaluation of this IC
is independent of our ideas about
QCD dynamics of hard interaction since it is based only on the data on
diffractive dissociation. Although this correction is a part of the
total CT pattern, it survives any modifications of the underlying dynamics and
should be considered as a {\it baseline} for CT studies.
 One can reliably disentangle this
contribution and the real CT effect only at $Q^2$ of a few tens of $GeV^2$, where
the former saturates, but CT provides a growth of nuclear
transparency up to unity.

\end{document}